\documentclass[%
reprint,
amsmath,amssymb,
aps,
pra,
]{revtex4-1}
\usepackage{graphicx}% Include figure files
\usepackage{dcolumn}% Align table columns on decimal point
\begin{document}
%\begin{CJK*}{GBK}{song}
%\preprint{APS/123-QED}

\title {
    % Rovibrational transition of HD determined to sub-ppb accuracy from Lamb-dip measurement \\
    Toward a determination of the proton-electron mass ratio from
     the Lamb-dip measurement of HD
    }
	
\author{
L.-G. Tao$^{1}$, A.-W. Liu$^{1,2}$, K. Pachucki$^{3}$, J. Komasa$^{4}$,
 Y. R. Sun$^{1,2}$, J. Wang$^{1}$, S.-M. Hu$^{1,2}$}
\email{Corresponding author. smhu@ustc.edu.cn}

\affiliation {
$^1$ Hefei National Laboratory for Physical Sciences at Microscale, $i$Chem center,
        University of Science and Technology of China, Hefei, 230026 China; \\
$^2$ CAS Center for Excellence %and Synergetic Innovation Center
 in Quantum Information and Quantum Physics,
        University of Science and Technology of China, Hefei, 230026 China; \\
$^{3}$ Faculty of Physics, University of Warsaw, Pasteura 5, 02-093, Warsaw, Poland; \\
$^{4}$ Faculty of Chemistry, Adam Mickiewicz University, Umultowska 89b, 61-614 Pozna\'{n}, Poland
}
\begin{abstract}
Precision spectroscopy of the hydrogen molecule is a test ground of
 quantum electrodynamics (QED), %physics beyond the Standard Model,
 and may serve for determination of fundamental constants.
Using a comb-locked cavity ring-down spectrometer,
 for the first time,
 we observed the Lamb-dip spectrum of the R(1) line in the overtone of HD. % at 1.4~$\mu$m,
%  which has a dipole transition rate of $2.1\times 10^{-5}$~s$^{-1}$.
The line position was determined to be  217 105 182.79 $\pm0.03_{stat}\pm0.08_{syst}$~MHz
 ($\delta\nu/\nu=4\times 10^{-10}$),
  which is the most accurate transition ever measured for the hydrogen molecule.
Moreover, from calculations including QED effects up to the order $m_e\alpha^6$,
 we obtained predictions for this R(1) line as well as for the HD dissociation energy,
 which are less accurate but
 signaling the importance of the complete treatment of nonadiabatic effects.
Provided that the theoretical calculation reaches the same accuracy,
 the present measurement will lead to a determination of the proton-electron mass ratio
  with a precision of 1.3 parts per billion.
\end{abstract}

%\pacs{06.20.Jr, 06.20.fb, 33.20.Ea}% PACS, the Physics and Astronomy
%33.15.Mt	Rotation, vibration, and vibration-rotation constants
%33.20.Ea	Infrared spectra
%06.20.fb	Standards and calibration
%06.20.Jr	Determination of fundamental constants
%06.30.Ft	Time and frequency

%\keywords{precision spectroscopy, metrology, cavity ring-down, CO}

\maketitle
%\end{CJK*}

%\section{Introduction}
%The hydrogen molecule is a prototype molecule in the development of quantum chemistry theories.
H$_2$, H$_2^+$, and their deuterated isotopologues are the simplest molecules
 whose energy levels can be derived from the quantum electrodynamics (QED) theory
 using a few fundamental physical constants:
  the Rydberg constant, the fine structure constant,
  the proton(deuteron)-electron mass ratio,
  and the proton(deuteron) charge radius.
The precision spectroscopy of molecular hydrogen has long been
 a test ground of the molecular theory~\cite{James1933JCP_H2,Wing1976PRL_HD+} and QED~\cite{Salumbides2011PRL_H2_QED,Korobov2017PRL_H2+}.
Comparison of the experimental and theoretical energy levels of molecular hydrogen
 also sets constraints on some hypotheses beyond the Standard Model,
 such as the long-distance fifth force between two hadrons~\cite{Salumbides2013PRD_5thForce}.
Having many long-lived rovibrational energy levels in the ground electronic states,
 the molecular hydrogen ion has been considered as a candidate for an optical clock~\cite{Schiller2014PRL_HDion_Clock}.
Recently, an agreement at 1~ppb (part per billion) accuracy between the experimental measurements
 and theoretical calculations has been demonstrated for HD$^+$,
 which allows for a determination of the proton-electron mass ratio
  with an accuracy of 2.9~ppb~\cite{Biesheuvel2016NC_HDion}.

It is more challenging to precisely calculate the energy levels of
 the four-body neutral hydrogen molecule than the three-body molecular hydrogen ion.
In the last half century,
 the accuracy of calculations of H$_2$ (and its isotopologues) in the ground electronic state
 has been continuously improved~\cite{Kolos1968JCP_H2_Calc, Wolniewicz1998AstroJSuppSer_H2_Calc, Pachucki2009JCP_H2_Calc, Pachucki2010PCCP_HD_Calc, Puchalski2016PRL_H2},
 and a precision of $10^{-6}$~cm$^{-1}$ ($10^4$~Hz) will be achievable in the near future~\cite{Pachucki2016JCP_H2,Puchalski2017PRA_H2}.
If the rovibrational transition frequencies of the hydrogen molecule are also measured with corresponding accuracy,
 it will lead to an improved determination of the proton-electron mass ratio $\mu_p\equiv m_p/m_e$.
The present $\mu_p$ value recommended by 2014 CODATA~\cite{CODATA2014RMP}
 has an uncertainty of 0.095~ppb.
However, a deviation of $3\sigma$ was observed by a recent measurement
 of the atomic mass of the proton~\cite{Heisse2017PRL_H_mass},
 indicating that more measurements from various methods with comparable uncertainties
  are needed for a consistency check of the constant.
%Till now, the best determination of $\mu_p$ using molecular energies is
% the laser spectroscopy of the HD$^+$ ion,
% which yielded a value of $\mu_p$ with an uncertainty of 2.9~ppb~\cite{Biesheuvel2016NC_HDion}.

In the electronic ground state,
 symmetric H$_2$ and D$_2$ molecules have only extremely weak quadrupole (E2) transitions,
 while HD exhibits weak dipole (E1) transitions due to nonadiabatic effects.
 % which was first predicted by Wick.
Although extensive spectroscopy of molecular hydrogen has been carried out
 (\cite{Campargue2012PCCP_H2_V2} and references therein)
  since the pioneering work by Herzberg in 1949~\cite{Herzberg1949Nature_H2},
 only Doppler-broadened spectra of the hydrogen molecule have been reported.
Attempts to improve the accuracy using the Doppler-limited spectra
 have been carried out for a few lines~\cite{Cheng2012PRA_H2,Mondelain2016JMS_D2_S2_V2},
 but the ambiguity in the line profile model
  may result in an uncertainty of several MHz~\cite{Wcislo2016PRA_H2}.
Sub-MHz accuracy is only possible when the line shape has been carefully investigated.
Doppler-free spectroscopy of the rovibrational transitions of molecular hydrogen
 is hindered by the very small transition rates.

Here we present the first Lamb-dip measurement of a rovibrational transition of molecular hydrogen.
The R(1) line in the $v=2-0$ band of HD has
 an Einstein coefficient of $2.1\times 10^{-5}$~s$^{-1}$~\cite{HITRAN2016},
 corresponding to a typical saturation power~\cite{Giusfredi2010PRL_CRDS_CO2_Sat}
  of $10^7$~W~cm$^{-2}$ at room temperature.
Taking the advantage of a high-finesse resonant cavity,
 we carried out saturation spectroscopy measurements using a continuous-wave diode laser
  with an output power of only several tens of milli-Watts.
%To the best of our knowledge, it is the weakest molecular line observed by saturation spectroscopy.
A sub-MHz line width was observed and the line center was determined
 %to be 217 105 182.79(9)~MHz
 with a fractional uncertainty of $4\times 10^{-10}$.
Compared with the previous value obtained from Doppler-limited spectra~\cite{Kassi2011JMS_HD_V2},
 the accuracy has been improved by a factor of 300.
This accuracy is so far the best among the experimental results
 of the hydrogen molecule including molecular hydrogen ions~\cite{Biesheuvel2016NC_HDion}.

\begin{figure}[htp]
	\centering
	\includegraphics[width=3.4in]{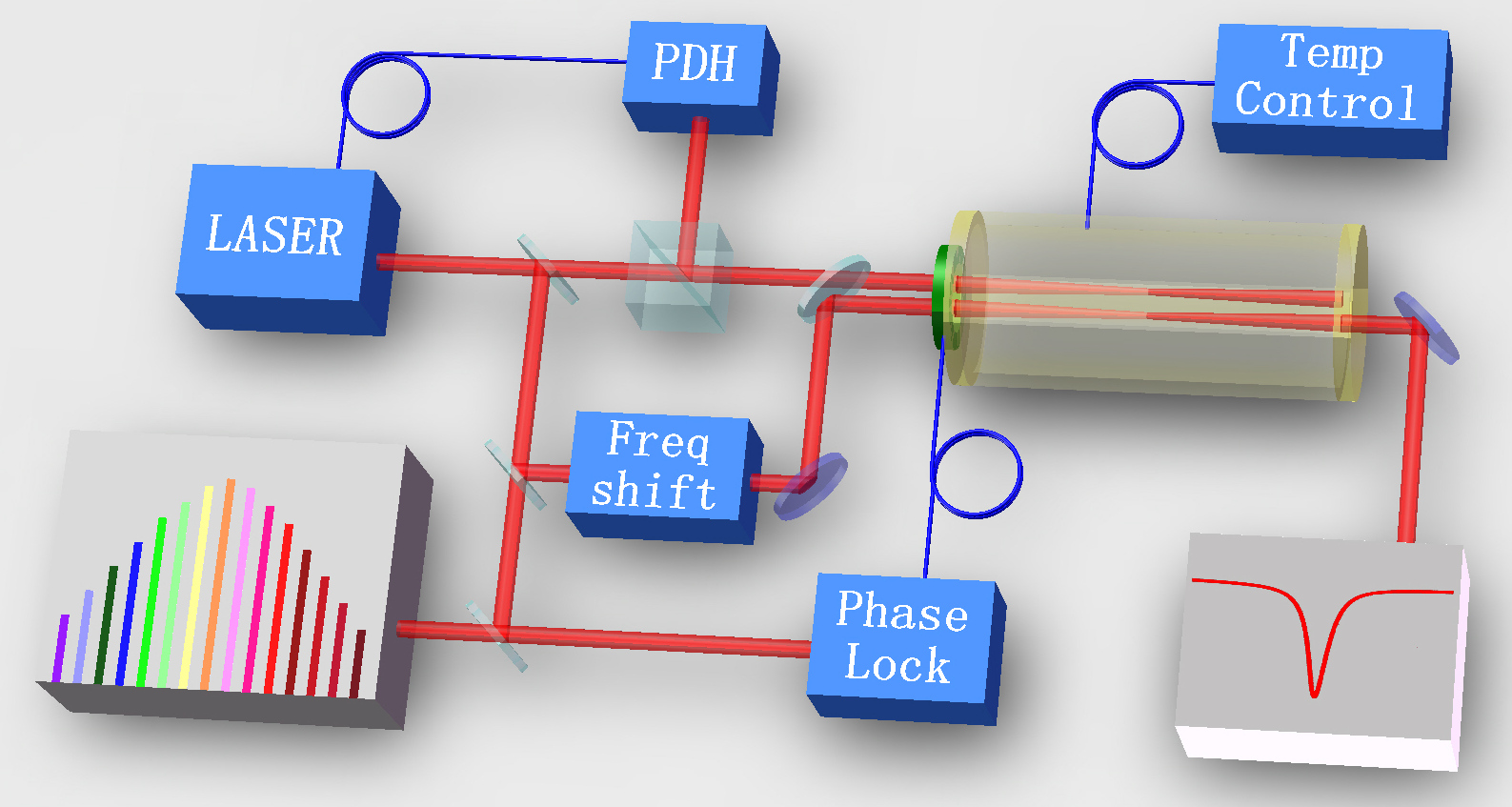}%CLRDS}
	\caption{ Configuration of the experimental setup.
        The probe laser frequency is locked with the cavity.
        Another beam from the probe laser
         is frequency shifted and used for CRDS measurement.
        Note the two beams are displaced in the figure for better illustration,
         but they are actually overlapped with each other in the cavity.
        The ring-down cavity length is locked according to the beat signal between
         the probe and a frequency comb.
		\label{Fig_Setup}
	}
\end{figure}

%\section{experimental}
The experimental setup is close to the one used in our previous study~\cite{Wang2017JCP_CO},
 and a diagram is presented in Fig.~\ref{Fig_Setup}.
An external-cavity diode laser is used as the probe laser,
 being locked to a ring-down (RD) cavity using the Pound-Drever-Hall (PDH) method.
The RD cavity is composed of a pair of high-reflectivity (HR) mirrors ($R=99.998$\%),
 corresponding to a finesse of $1.2\times 10^5$.
The 80~cm-long RD cavity is temperature stabilized at 25~$^\circ$C
 and the fluctuation is below 10~mK.
The cavity length is stabilized through a piezo actuator (PZT)
 by a phase-lock circuit driving by the beat signal
 between the probe laser and an optical frequency comb.
The frequency comb is synthesized by an Er:fiber oscillator operated at 1.56~$\mu$m.
Its repetition frequency ($f_R\approx 200$~MHz) and carrier offset frequency ($f_0$)
 are both referenced to a GPS-disciplined rubidium clock (SRS FS725).
%%%%%%%%%%%%%
A separated beam from the probe laser,
 frequency shifted by an acousto-optic modulator (AOM)
  and an electro-optic modulator (EOM),
 is coupled into the RD cavity from another side of the cavity.
The frequency shift is set exactly as the difference between two longitudinal modes
 of the ring-down cavity.
The AOM also serves as a beam chopper to initiate the ring-down signal.
The ring-down curve is fit by an exponential decay function,
 and the sample absorption coefficient $\alpha$ is determined by:
 %~\cite{Zalicki1995JCP_CRDS_Quantitative}:
  $\alpha = (c\tau)^{-1} - (c\tau_0)^{-1}$,
 where $c$ is the speed of light,
  and $\tau$ and $\tau_0$ are the decay times of the cavity
   with and without sample, respectively.

\begin{figure}[htp]
	\centering
	\includegraphics[width=3.4in]{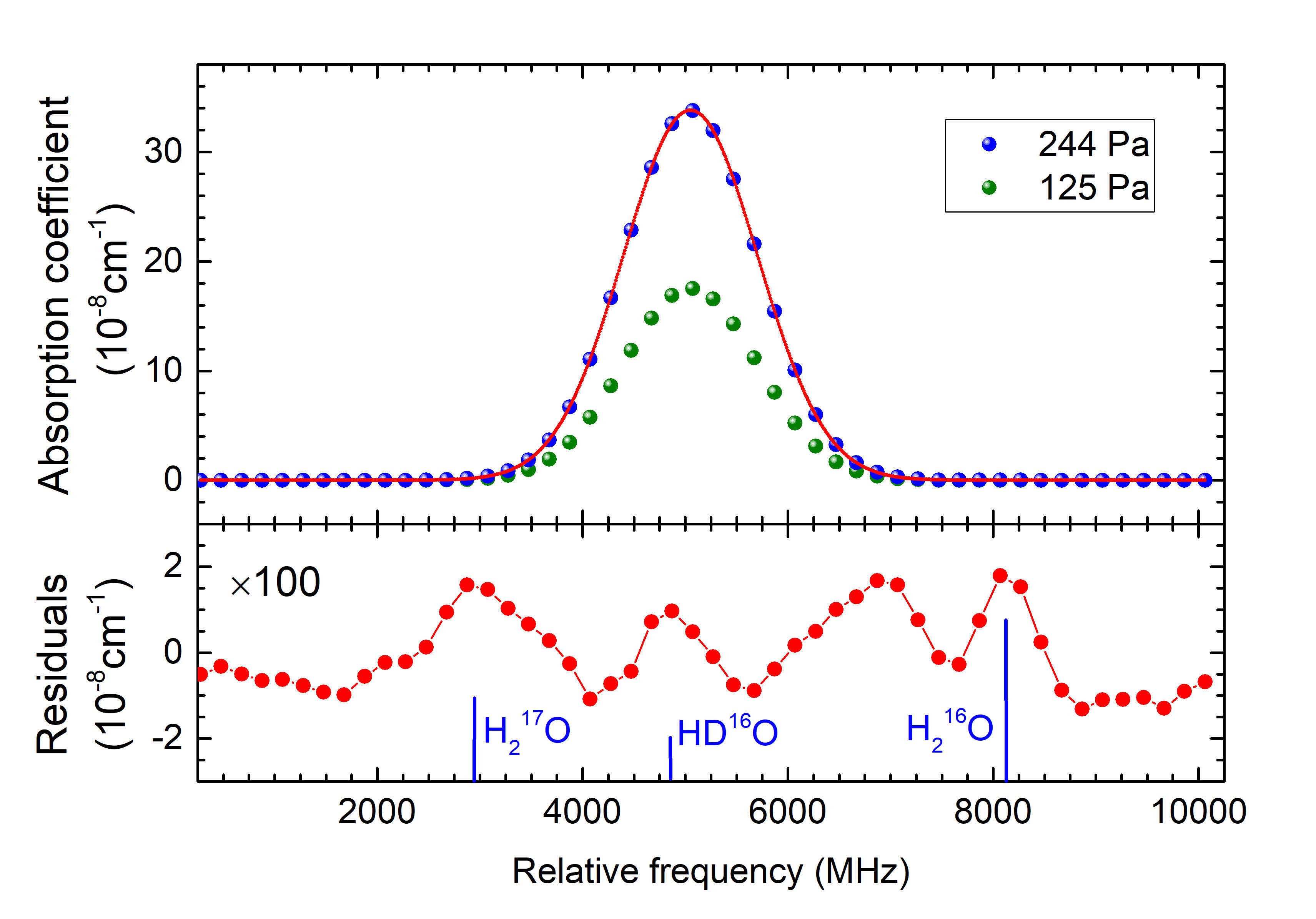}
	\caption{
    Doppler-broadened cavity ring-down spectra of the R(1) line in the 2-0 band of HD.
    The lower panel shows fitting residuals of the spectrum recorded at 244~Pa
     using a Gaussian function.
    The positions of a few weak water lines in the vicinity of the spectrum are also
     marked on the figure.
    The water line positions and relative intensities are according to the values given in the HITRAN database~\cite{HITRAN2016}.
    	\label{Fig-Doppler}
	}
\end{figure}

The R(1) line in the $2-0$ overtone band of HD
 is located at 7241.85~cm$^{-1}$,
 and the line intensity is $3.6\times 10^{-25}$cm/molecule~\cite{Kassi2011JMS_HD_V2}.
The HD sample was purchased from Sigma-Aldrich Co. and used without further purification.
The Doppler-broadened spectrum recorded at 125~Pa and 244~Pa are
  shown in Fig.~\ref{Fig-Doppler}.
By fitting the spectrum with a Gaussian function,
 we derived a line center of 217 105 181(2)~MHz and
   a Gaussian width (half width at half maximum, HWHM) of 771~MHz.
The Gaussian width agrees well with the calculated Doppler width of 775~MHz at 298~K.
The uncertainty of the line position mainly comes from
 the parasitic optical interference (``fringes''),
 the collision effect~\cite{Wcislo2016PRA_H2},
 and the influence due to a few nearby water absorption lines
 which presented as trace contamination in the ring-down cavity.

\begin{figure}[htp]
	\centering
	\includegraphics[width=3.4in]{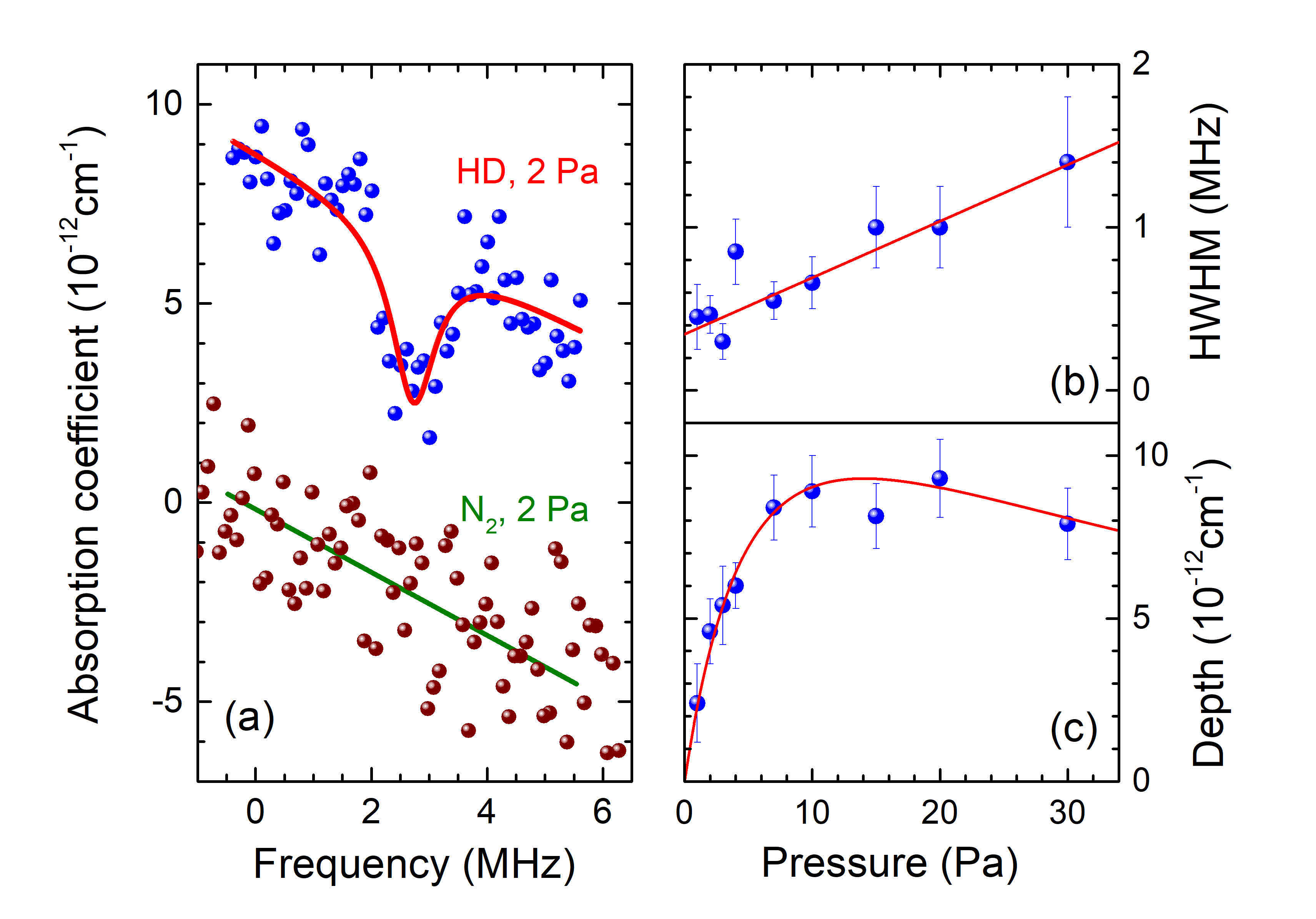}
	\caption{
    Cavity ring-down spectra at 7241.8494~cm$^{-1}$
     recorded with HD sample (a, upper) and pure nitrogen (a, lower).
    Lamb dip of the R(1) line was fit with a Lorentzian function.
    The width (HWHM, half width at half maximum) (b) and depth (c)
    of the Lamb dip vary with the sample pressure of HD.
    	\label{Fig-spec}
	}
\end{figure}

Sample pressures below 30~Pa were used for Lamb-dip measurements.
The laser power used for spectral probing was about 15~mW
 and the intra-cavity laser power was estimated~\cite{Ma1999JOSAB_C2H2,Wang2017RSI}
  to be about 200~W, %3E4W/cm2
 leading to a saturation parameter of about 0.2\% (maximum)
  with a laser beam waist radius of 0.5~mm.
The spectrum recorded at a pressure of 2~Pa is shown in Fig.~\ref{Fig-spec}(a).
It is an average of about 400~scans
 taken from a continuous measurement of about 12 hours.
The Lamb dip of the R(1) line has a width (HWHM) of about 0.4~MHz
  and a depth of about $5\times 10^{-12}$~cm$^{-1}$.
For comparison, a spectrum recorded with pure nitrogen gas is also given in the same figure.

The Lamb-dip central position, width, and depth
 were derived from a fit of the spectrum using a Lorentzian function.
As shown in Fig.~\ref{Fig-spec}(b),
 the depth and width of the Lamb dip of the R(1) line
 vary with the sample pressures,
 and they can be well described by the collision-induced broadening effect.
The line width is mainly due to the transit-time broadening (0.35~MHz)
 and the collision-induced broadening (0.03~MHz~Pa$^{-1}$).
The depth of the Lamb-dip is proportional to the coefficient:
 $D\propto (1+S)^{-1/2} - (1+2S)^{-1/2}$,
 where $S$ is the saturation parameter.
%%%%%%%%% In lower panel, using collision-induced broadening of 0.06~MHz~Pa$^{-1}$  and Einstein coefficient of $2.1\times 10^{-5}$~s$^{-1}$, the slmulation of the depth at different pressure(red line) fits well with the expriment(blue dots).%%%%%%%%%%%%%

\begin{figure}[htp]
	\centering
	\includegraphics[width=3in]{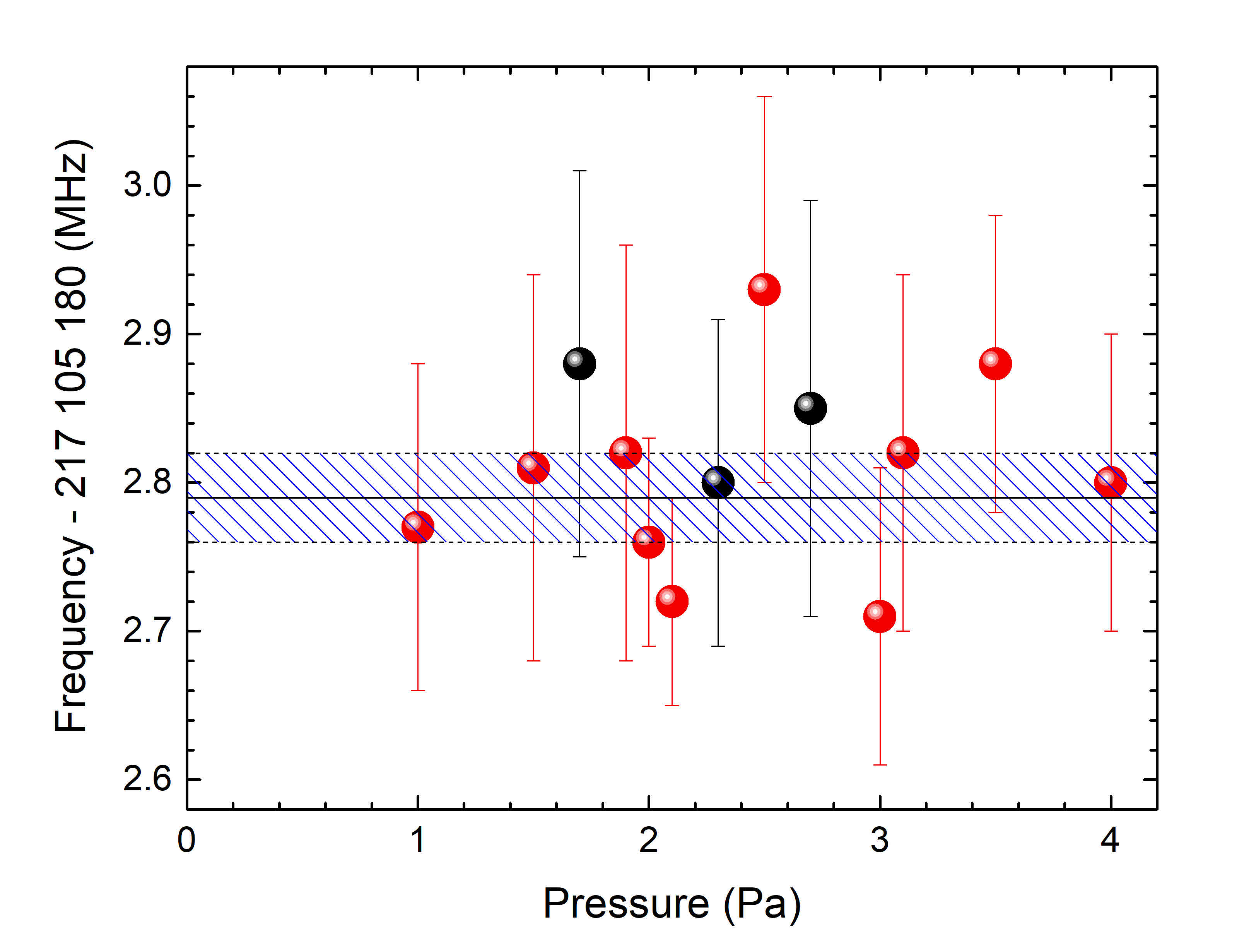}
	\caption{
    R(1) positions determined from spectra recorded at HD sample pressures of 1-4~Pa.
    Black and red points indicates measurements by switching the laser beams used
    for frequency locking and spectral probing.
    The region of shadow indicates the average value with $1\sigma$ uncertainty.
		\label{Fig-position}
	}
\end{figure}

In order to reduce the influence due to the collision-induced shift,
 the line center was determined from the spectra recorded with
  HD sample pressures of 1 - 4~Pa,
 as shown in Fig.~\ref{Fig-position}.
In this pressure region, no evidence of the pressure-induced shift has been observed.
A statistical uncertainty of 0.03~MHz
 was obtained from 2600 scans recorded in 87 hours.
A major systematic uncertainty arises from the possible asymmetry in the line profile
 which would lead to a bias on the line center derived from the fit of the spectrum.
We have examined the low-pressure spectra
 and concluded that such asymmetry, if any, should be below the present noise level.
Taking a signal-to-noise ratio of 5:1 and a line width (HWHM) of 0.4~MHz,
 we give an up-limit uncertainty of 0.08~MHz due to the line profile model.
Other contributions to the uncertainty budget are much smaller and negligible in this study.
The laser frequency is calibrated by the frequency comb
 and eventually by the GPS-disciplined rubidium clock
 which has a fractional uncertainty of $2\times 10^{-12}$ (0.4~kHz at 1.4~$\mu$m).
The radio frequencies used to drive the AOM and EOM have a drift below 50~Hz.
The second-order Doppler shift is 3~kHz.
%The recoil shift is -34.7~kHz.
Note that beside the spectral beam which is on resonance with the transition,
 another laser beam used to lock the laser frequency,
  being on resonance with a nearby cavity mode,
 also presents in the cavity.
We purposely switched between the two beams and repeated the measurement,
 %equivalent to use a negative $f_{AOM}$,
 but found no difference (black and red points in Fig.~\ref{Fig-position})
  within the experimental uncertainty.
%%%%%%%%%The AC Strak shfit if negligible because of the  weak diople moment and sparse rovibrational energy level.%%%%%%%%%%
The final value of the line position determined in this work is:
\begin{eqnarray}
\nu_0 &=& 217\, 105\, 182.79(3)_{stat.}(8)_{syst.}~{\rm MHz} \nonumber \\
  &=& 7\, 241.849\, 386(1)_{stat.}(3)_{syst.} {\rm cm}^{-1}
  \label{EQ-nu}
\end{eqnarray}
The R(1) line frequency determined in this study agrees with the value 7241.8497(10)~cm$^{-1}$
 derived from Doppler-limited spectra reported by Kassi \textit{et al.}~\cite{Kassi2011JMS_HD_V2},
 while the accuracy has been improved by a factor of 300.

\begin{table}[htp]
\centering
\caption{Calculated and experimental energies of HD (unit: cm$^{-1}$).
There is an implicit relative uncertainty of about $8\times 10^{-4}$ in
   $E^{(4)}$ and $E^{(5)}$ due to nonadiabatic corrections.
\label{Table}
}
\begin{tabular}{cr@{.}lr@{.}l}
\hline \hline
  & \multicolumn{2}{c}{$D_0$, (0,0)} & \multicolumn{2}{c}{2-0, R(1)} \\
\hline
$E^{(2)}$   & 36406&510839(1)     & 7241&846169(1)     \\
$E^{(4)}$   &    -0&531325(1)     &    0&040719        \\
$E^{(5)}$   &    -0&1964(2)       &   -0&03743(4)      \\
$E^{(6)}$   &    -0&002080(6)     &   -0&000339        \\
$E^{(7)}$   &     0&00012(6)      &    0&000021        \\
$E_{FS} $   &    -0&000117        &   -0&000021        \\
Total       & 36405&7810(5)       & 7241&84912(6)      \\
\hline
%Expt.       & 36405&78366(36)$^a$ & 7241&8497(10)$^b$  \\ %& 7361&9037(10)$^b$\\
Expt.       & 36405&78366(36)$^a$ & 7241&849386(3)     \\
Diff.       &     0&0026          &    0&00027         \\
\hline \hline
\end{tabular}
\flushleft\noindent
$^a$ From Ref.~\cite{Sprecher2010JCP_HD}. \\
%$^b$ From Ref.~\cite{Kassi2011JMS_HD_V2}.
\end{table}

Our theoretical value, as given in Table~\ref{Table}, amounts to $7241.849\,12$(6)~cm$^{-1}$.
It was obtained as follows.
The energy of a rovibrational level is expanded
 in powers of the fine-structure constant $\alpha$:
 \begin{equation}
 \label{EQ-E}
 E = \sum^\infty_{n=2}E^{(n)}
 \end{equation}
 where each $E^{(n)}$ is proportional to $\alpha^n$ (and may contain $\ln\alpha$).
The leading term of this expansion, $E^{(2)}$ is the non-relativistic energy.
It was calculated without any approximations,
 using nonadiabatic explicitly correlated wave function,
 with a numerical uncertainty of $10^{-6}$~cm$^{-1}$.
This is the part that has been significantly improved
 with respect to previous studies~\cite{Pachucki2010PCCP_HD_Calc}.
Other expansion terms in Eq.~(\ref{EQ-E}) were calculated within the adiabatic approximation.
The next term $E^{(3)}$ is absent,
 $E^{(4)}$ is the relativistic correction~\cite{Puchalski2017PRA_H2},
 $E^{(5)}$ is the QED correction~\cite{Piszczatowski2009JCTC_H2},
 and the terms with $n\geq 6$ constitute higher-order relativistic and QED corrections.
The recent accurate calculation of $E^{(6)}$~\cite{Puchalski2016PRL_H2}
 was a significant step in achieving high-precision theoretical predictions.
Although numerical uncertainties in $E^{(n)}$ are at the order of $10^{-6}$~cm$^{-1}$ or less,
 as shown in Table~\ref{Table},
 the discrepancies with experiment for the dissociation energy~\cite{Sprecher2010JCP_HD}
  and the R(1) transition are 0.0026~cm$^{-1}$ and 0.00027~cm$^{-1}$, respectively.
They are both about five times the estimated theoretical uncertainty given here.
The most probable reason is underestimation of relativistic nonadiabatic effects.
A preliminary estimate of these effects is $E^{(4)}$ times the electron-nuclear mass ratio,
 which is about 10 times smaller than the discrepancy.

In the calculation, we used the CODATA recommended values~\cite{CODATA2014RMP} of
 the following constants:
 the Rydberg constant $R_y=109\,737.315\,685\,08(65)$~cm$^{-1}$,
 the fine-structure constant $\alpha  =0.007\,297\,352\,566\,4(17)$,
 and the proton/deuteron-electron mass ratios
 $\mu_p =    1836.152\,673\,89(17)$,
 $\mu_d \equiv m_d/m_e =    3670.482\,967\,85(13)$.
For the proton and deuteron charge radii, we used the values from
 the muonic hydrogen measurements~\cite{Antognini2013Science_H_Mu}:
 $r_p = 0.840087(39)$~fm and $r_d = 2.12771(22)$~fm.
The deviation in the HD transition frequency $\nu$
 can be translated to deviations of the physical constants:
\begin{equation} \label{EQ-dmu}
\frac{d\nu}{\nu} =
  \beta_{R_y} \frac{d R_y}{R_y} + \beta_{\alpha} \frac{d\alpha}{\alpha}
 + \beta_{\mu_p} \frac{d\mu_p}{\mu_p} + \beta_{\mu_d} \frac{d\mu_d}{\mu_d}
 + \beta_{r^2} \frac{d r^2}{r^2}
\end{equation}
where $r^2= r_p^2+ r_d^2$ is
 the sum of the nuclear charge radii squares of proton and deuteron.
For the 2-0 R(1) transition of HD,
 the $\beta$ coefficients are as follows:
 $\beta_{R_y}=1$, $\beta_{\alpha}=-4.3\times 10^{-6}$,
  $\beta_{\mu_p}=-0.31$, $\beta_{\mu_d} = -0.060$,
  and $\beta_{r^2}=-2.9\times 10^{-9}$. %assuming the deuteron-proton mass ratio is fixed.
Taking into account the relative uncertainties of these constants,
 the most significant term in Eq.~(\ref{EQ-dmu}) is $\beta_{\mu_p}\frac{d\mu_p}{\mu_p}$.
Therefore, the transition frequency measured in this work could
 lead to a determination of the $\mu_p$ value with an uncertainty of 1.3~ppb
 if the theoretical calculation reaches the corresponding precision.

Note that the current experimental accuracy is mainly limited by the line width due to
 transit-time broadening.
The accuracy could be considerably improved by conducting
 cavity-enhanced saturation spectroscopy of sample gases cooled to a few Kelvin
  by buffer-gas cooling~\cite{Spaum2016Nat_Comb_CES_Buffer}.
In this case, the width of the Lamb dip would decrease by an order of magnitude.
Moreover, the reduced line width will also reduce the saturation power of the transition
 and result in improved signal-to-noise ratio in the Lamb-dip measurement.
As a result, a fractional accuracy of $10^{-12}$ of the HD transition frequency
  is expected.
On the theoretical side, it has been recently demonstrated that
 the numerical solution of the Schr\"{o}dinger equation of molecular hydrogen
 can be as accurate as $10^{-12}$,
 which paves the way for using the energy levels of molecular hydrogen to determine
  other physical constants in Eq.~(\ref{EQ-dmu}), such as
 the Rydberg constant and the proton charge radius~\cite{Puchalski2016PRL_H2,Puchalski2017PRA_H2},
  similar to their determination from the spectroscopy of atomic hydrogen~\cite{Beyer2017Science_H}.

{\, }

%\section*{acknowledgments}
This work was jointly supported by the Chines Academy of Science (XDB21020100)
 and the National Natural Science Foundation of China (21688102, 91436209, 21427804).
The computational part of this project was supported by NCN grant 2017/25/B/ST4/01024
 and by Poznan Supercomputing and Networking Center.

%\bibliography{H2}


\end{document}